\pdfoutput=1
\RequirePackage{ifpdf}
\ifpdf 
\documentclass[pdftex]{sigma}
\else
\documentclass{sigma}
\fi

\numberwithin{equation}{section}

\usepackage{cite}

\begin{document}

\allowdisplaybreaks

\renewcommand{\PaperNumber}{080}

\FirstPageHeading

\ShortArticleName{Enlarged Shape Invariance}

\ArticleName{Novel Enlarged Shape Invariance Property\\ and Exactly Solvable Rational Extensions\\ of the Rosen--Morse II and Eckart Potentials}

\Author{Christiane QUESNE}

\AuthorNameForHeading{C.~Quesne}

\Address{Physique Nucl\'eaire Th\'eorique et Physique Math\'ematique,  Universit\'e Libre de
Bruxelles, \\ Campus de la Plaine CP229, Boulevard~du Triomphe, B-1050 Brussels,
Belgium}
\Email{\href{mailto:cquesne@ulb.ac.be}{cquesne@ulb.ac.be}}

\ArticleDates{Received August 30, 2012, in f\/inal form October 15, 2012; Published online October 26, 2012}

\Abstract{The existence of a novel enlarged shape invariance property valid for some rational extensions of shape-invariant conventional potentials, f\/irst pointed out in the case of the Morse potential, is conf\/irmed by deriving all rational extensions of the Rosen--Morse II and Eckart potentials that can be obtained in f\/irst-order supersymmetric quantum mechanics. Such extensions are shown to belong to three dif\/ferent types, the f\/irst two strictly isospectral to some starting conventional potential with dif\/ferent parameters and the third with an extra bound state below the spectrum of the latter. In the isospectral cases, the partner of the rational extensions resulting from the deletion of their ground state can be obtained by translating both the potential parameter $A$ (as in the conventional case) and the degree $m$ of the polynomial arising in the denominator. It therefore belongs to the same family of extensions, which turns out to be closed.}


\Classification{81Q05; 81Q60}

\vspace{-2mm}

\section{Introduction}

During the last few years, a lot of research activity has been devoted to the construction of new exactly solvable rational extensions of well-known quantum potentials, some of which are connected with the novel f\/ield of exceptional orthogonal polynomials (EOP) and with the appearance of a so far unsuspected class of (translationally) shape-invariant (SI) potentials \cite{gomez04a, gomez04b, gomez09, gomez10a, gomez10b, gomez12a, gomez12b, gomez12c, cq08, bagchi, cq09, cq11a, cq11b, cq11c, cq12, odake09, odake10a, odake10b, ho11a, sasaki10, odake11, sasaki12, ho11b, ho11c, grandati11a, grandati11b, grandati11c, grandati12, dutta10, berger, dutta11, bougie10, bougie11, ramos, carinena, fellows}.

{\sloppy Among several equivalent approaches, such as supersymmetric quantum mechanics (SUSYQM), Darboux--Crum transformations, Darboux--B\"acklund ones, and prepotential method, we have resorted from the very beginning to the f\/irst procedure either in its (standard) f\/irst-order form \cite{sukumar, cooper} or in its higher-order one \cite{andrianov, bagrov, aoyama, fernandez}. In the former model, one starts from some nodeless solution~$\phi(x)$ of a SI potential~$V^{(+)}(x)$ Schr\"odinger equation, corresponding to an energy eigenvalue~$E$ less than or equal to the ground-state energy~$E^{(+)}_0$. From the factorization function $\phi(x)$, one then constructs the so-called partner potential~$V^{(-)}(x)$. Whenever the factorization function $E$ is smaller than $E^{(+)}_0$ and $\phi(x)$ is chosen of polynomial type, $V^{(-)}(x)$ turns out to be an algebraic deformation \cite{gomez04a} or rational extension of a potential similar to $V^{(+)}(x)$, but with dif\/ferent parameters~\cite{bagchi, cq09}. According to whether $\phi^{-1}(x)$ is norma\-li\-zab\-le or not, $V^{(-)}(x)$ has an additional bound state below the spectrum of~$V^{(+)}(x)$ (unbroken SUSYQM) or both potentials are strictly isospectral (broken SUSYQM).

}

For the radial oscillator, Scarf I, and generalized P\"oschl--Teller potentials (or, equivalently, isotonic oscillator, trigonometric and hyperbolic P\"oschl--Teller potentials), the bound-state wavefunctions of $V^{(-)}(x)$ can be expressed in terms of EOP\footnote{Only a f\/inite number of such polynomials are found in the case of the generalized (or hyperbolic) P\"oschl--Teller potential, which has a f\/inite bound-state spectrum.} and, in the broken SUSYQM case, the SI property of $V^{(+)}(x)$ is preserved when going to its partner.

In a recent work \cite{cq12}, we have constructed rational extensions of the Morse potential $V_{A,B}(x)$ in such a framework and shown that in contrast with what happens for the above-mentioned potentials, the extended potentials obtained in  the broken SUSYQM case do not have the SI property of the Morse potential. Nevertheless, they exhibit an unfamiliar extended SI property, in the sense that their partner is obtained by translating both the potential parameter $A$ (as in the conventional case) and the degree $m$ of the polynomial arising in the denominator, and therefore belongs to the same family of extended potentials.

The aim of the present paper is to uncover other classes of rationally-extended potentials displaying such a novel enlarged SI property. For such a purpose, we plan to start from some SI potentials whose bound-state wavefunctions can be expressed in terms of Jacobi polynomials.

In Section~\ref{section2}, we review the case of the Rosen--Morse II potential (also termed hyperbolic Rosen--Morse potential). In Section~\ref{section3}, a similar study is carried out for the Eckart potential. Section~\ref{section4} deals with the enlarged SI property of the extended potentials that are obtained in broken SUSYQM. Finally, Section~\ref{section5} contains the conclusion.

\section[Rationally-extended Rosen-Morse II potentials in first-order SUSYQM]{Rationally-extended Rosen--Morse II potentials\\ in f\/irst-order SUSYQM}\label{section2}

\subsection{General results}\label{section2.1}

The Rosen--Morse II potential
\begin{gather*}
  V_{A,B}(x) = - A(A+1) \operatorname{sech}^2 x + 2B \tanh x, \qquad - \infty < x < \infty,
\end{gather*}
where we assume $A > 0$ and $0 < B < A^2$,\footnote{In this paper, we take units wherein $\hbar = 2m = 1$. The parameter $B$ is assumed positive for convenience since changing $B$ into $-B$ only amounts to changing $x$ into $-x$. In contrast, the hypotheses $A > 0$ and $|B| < A^2$ are necessary for getting at least one bound state.} is known to have a f\/inite number of bound states, whose energy and wavefunction are given by (see, e.g.,~\cite{cooper})
\begin{gather*}
  E^{(A,B)}_{\nu} = - (A - \nu)^2 - \frac{B^2}{(A - \nu)^2}, \quad \nu = 0, 1, \ldots, \nu_{\max}, \!\qquad
  A - 1 - \sqrt{B} \le \nu_{\max} < A - \sqrt{B},
\end{gather*}
and
\begin{gather*}
  \psi^{(A,B)}_{\nu}(x)  \propto (\operatorname{sech}x)^{A-\nu} \exp \left(- \frac{B}{A-\nu} x\right)
      P_{\nu}^{\left(A-\nu+\frac{B}{A-\nu}, A-\nu-\frac{B}{A-\nu}\right)}(\tanh x) \\
\hphantom{\psi^{(A,B)}_{\nu}(x) }{}
\propto (1-z)^{\frac{1}{2}\left(A-\nu+\frac{B}{A-\nu}\right)} (1+z)^{\frac{1}{2}\left(A-\nu-\frac{B}{A-\nu}
      \right)} P_{\nu}^{\left(A-\nu+\frac{B}{A-\nu}, A-\nu-\frac{B}{A-\nu}\right)}(z),
\end{gather*}
respectively. Here
\begin{gather}
  z = \tanh x, \qquad -1 < z < 1,  \label{eq:z}
\end{gather}
and $P_{\nu}^{\left(A-\nu+\frac{B}{A-\nu}, A-\nu-\frac{B}{A-\nu}\right)}(z)$ denotes a Jacobi polynomial.

To construct rational extensions of this potential, we have to determine all polynomial-type, nodeless solutions $\phi(x)$ of the Schr\"odinger equation
\begin{gather}
  \left(- \frac{d^2}{dx^2} + V_{A,B}(x)\right) \phi(x) = E \phi(x)  \label{eq:SE}
\end{gather}
with $E < E^{(A,B)}_0 = - A^2 - \frac{B^2}{A^2}$. In terms of the variable $z$ def\/ined in (\ref{eq:z}), equation (\ref{eq:SE}) can be rewritten as
\begin{gather}
  \left[- \big(1-z^2\big)^2 \frac{d^2}{dz^2} + 2z \big(1-z^2\big) \frac{d}{dz} - A(A+1) \big(1-z^2\big) + 2B z\right] \phi\bigl(x(z)\bigr)
  = E \phi\bigl(x(z)\bigr).  \label{eq:SE-bis}
\end{gather}

For such a purpose, let us make the changes of variable and of function
\begin{gather*}
  t = \tfrac{1}{2} (1-z), \qquad \phi\bigl(x(z)\bigr) = t^{\lambda} (1-t)^{\mu} f(t),
\end{gather*}
where $\lambda$, $\mu$, and $f(t)$ are two constants and a function, respectively. The resulting equation for $f(t)$ reduces to the hypergeometric equation
\begin{gather}
  \left\{t(1-t) \frac{d^2}{dt^2} + [c - (a+b+1)t] \frac{d}{dt} - ab\right\} f(t) = 0  \label{eq:hyper}
\end{gather}
provided the conditions
\begin{gather}
  E = 2B - 4\lambda^2, \qquad \lambda^2 - \mu^2 = B, \qquad a = \lambda + \mu - A, \nonumber\\
  b = \lambda +
  \mu + A + 1, \qquad c = 2\lambda + 1  \label{eq:cond}
\end{gather}
are satisf\/ied.

On restricting ourselves to the regular solution ${}_2F_1(a, b; c; t)$ of~(\ref{eq:hyper}), we f\/ind four polynomial-type solutions, expressed in terms of Jacobi polynomials, if and only if either $a$ or $c-a$ is an integer~\cite{erdelyi} (see also~\cite{gomez04a}),
\begin{gather}
  f_1(t)  = {}_2F_1(a, b; 1+a+b-c; 1-t) \propto P_m^{(b-c-m, c-1)}(2t-1) \qquad \text{for} \ a=-m, \nonumber\\
  f_2(t)  = t^{1-c} (1-t)^{c-a-b} {}_2F_1(1-a, 1-b; 1-a-b+c; 1-t) \nonumber\\
   \hphantom{f_2(t)}{}
  \propto t^{1-c} (1-t)^{c-b-1-m} P_m^{(c-b-1-m, 1-c)}(2t-1) \qquad \text{for}  \ a=m+1, \nonumber\\
  f_3(t)  = (1-t)^{c-a-b} {}_2F_1(c-a, c-b; 1-a-b+c; 1-t) \label{eq:reg-sol}\\
  \hphantom{f_3(t) }{}
   \propto (1-t)^{-b-m} P_m^{(-b-m, c-1)}(2t-1) \qquad \text{for} \ c-a=-m,\nonumber\\
  f_4(t)  = t^{1-c} {}_2F_1(a+1-c, b+1-c; 1+a+b-c; 1-t) \nonumber\\
  \hphantom{f_4(t)}{}
  \propto t^{1-c} P_m^{(b-1-m, 1-c)}(2t-1) \qquad \text{for} \ c-a=m+1.\nonumber
\end{gather}
Combining (\ref{eq:cond}) with the condition found for $a$ or $c-a$ in (\ref{eq:reg-sol}) leads to two independent polynomial-type solutions of (\ref{eq:SE-bis}),
\begin{gather}
  \phi_1(x)  = (1-z)^{\frac{1}{2}\left(A-m + \frac{B}{A-m}\right)} (1+z)^{\frac{1}{2}\left(A-m - \frac{B}{A-m}
      \right)} P_m^{\left(A-m + \frac{B}{A-m}, A-m - \frac{B}{A-m}\right)}(z), \nonumber\\
  E_1  = - (A-m)^2 - \frac{B^2}{(A-m)^2},
  \label{eq:phi-1}
\end{gather}
and
\begin{gather}
  \phi_2(x)  = (1-z)^{-\frac{1}{2}\left(A+m+1 + \frac{B}{A+m+1}\right)} (1+z)^{-\frac{1}{2}\left(A+m+1
      - \frac{B}{A+m+1}\right)} \nonumber\\
\hphantom{\phi_2(x)  =}{}  \times P_m^{\left(-A-m-1 - \frac{B}{A+m+1}, -A-m-1 + \frac{B}{A+m+1}\right)}(z), \nonumber\\
  E_2  = - (A+m+1)^2 - \frac{B^2}{(A+m+1)^2},
   \label{eq:phi-2}
\end{gather}
coming from $f_1(t)$ (or $f_4(t)$) and $f_2(t)$ (or $f_3(t)$), respectively.

For the f\/irst solution $\phi_1(x)$, the condition on the energy $E_1 < E_0^{(A,B)}$ is satisf\/ied if and only if the parameters $A$ and $B$ vary in anyone of the following three ranges:
\begin{alignat*}{3}
  & ({\rm 1a}) \quad &&  A > m, \quad A(A-m) < B < A^2; & \\
  & ({\rm 1b}) \quad && \tfrac{m}{2} < A < m, \quad - A(A-m) < B < A^2; & \\
  & ({\rm 1c}) \quad &&  0 < A < \tfrac{m}{2}, \quad 0 < B < A^2. &
\end{alignat*}
In contrast, for the second solution $\phi_2(x)$, the condition is fulf\/illed for all allowed $A$ and $B$ values, namely $A > 0$ and $0 < B < A^2$.

It only remains to check whether the Jacobi polynomial in (\ref{eq:phi-1}) or (\ref{eq:phi-2}) is free from any zero in the interval $(-1, +1)$. In Appendix~\ref{appendixA}, from the known distribution of the zeros of the general Jacobi polynomial $P_n^{(\alpha, \beta)}(x)$ on the real line~\cite{szego} (see also~\cite{erdelyi}), we formulate a convenient rule enumerating the cases where there is no zero in $(-1, +1)$ (Rule~1). For the f\/irst solution~$\phi_1(x)$, it can be readily shown that the parameters~$\alpha$, $\beta$ in~(\ref{eq:phi-1}) satisfy the conditions $\alpha > 0$, $\beta < -m$ in Case~1a, $\alpha < -m$, $\beta > 0$ in Case~1b, and $\alpha < \beta < 0$ together with $\beta > -m$ in Case~1c. The f\/irst two are therefore associated with Cases~a and~b of Rule~1, whereas the last one may correspond to some exceptional subcases of Case~c for appropriately chosen parameters. None is found for $m=1$, but for $m=2$, $0 < A < 1$, $0 < B < A^2$, for instance, there exists one for $A \ne \frac{1}{2}$ and $0 < B < \min\bigl(A^2, (1-A) (2-A)\bigr)$. For the second solution $\phi_2(x)$, the parame\-ters~$\alpha$,~$\beta$ in~(\ref{eq:phi-2}) fulf\/il the conditions $\alpha < -m$, $\beta < -m$, and therefore correspond to Case~c of Rule~1 (nonexceptional subcase) provided~$m$ is chosen even ($m=2k$).

We conclude that, apart from some exceptional cases, which we are going to omit for simplicity's sake\footnote{It is worth noting that some exceptional cases also exist for rationally-extended radial oscillator (or isotonic) potentials, but they were not considered in~\cite{grandati11a}.}, there exist three acceptable polynomial-type, nodeless solutions of the Rosen--Morse~II Schr\"odinger equation,
\begin{gather}
  \phi^{\rm I}_{A,B,m}(x)  = \chi^{\rm I}_{A,B,m}(z) P_m^{\left(A-m + \frac{B}{A-m}, A-m - \frac{B}{A-m}\right)}
       (z) \nonumber\\
\hphantom{\phi^{\rm I}_{A,B,m}(x)  =}{} \ \text{if} \ m=1, 2, 3, \ldots, \quad A > m, \quad A(A-m) < B < A^2,
  \label{eq:phi-I}
\\
  \phi^{\rm II}_{A,B,m}(x)  = \chi^{\rm II}_{A,B,m}(z) P_m^{\left(A-m + \frac{B}{A-m}, A-m - \frac{B}{A-m}\right)}
       (z) \nonumber\\
\hphantom{\phi^{\rm II}_{A,B,m}(x)  =}{}  \ \text{if} \ m=1, 2, 3, \ldots, \quad \tfrac{m}{2} < A < m, \quad -A(A-m) < B < A^2,
  \label{eq:phi-II}
\\
  \phi^{\rm III}_{A,B,m}(x)  = \chi^{\rm III}_{A,B,m}(z)
       P_m^{\left(-A-m-1 - \frac{B}{A+m+1}, -A-m-1 + \frac{B}{A+m+1}\right)}(z) \nonumber\\
\hphantom{\phi^{\rm III}_{A,B,m}(x)  =}{}
 \ \text{if} \ m=2, 4, 6, \ldots, \quad A > 0, \quad 0 < B < A^2,
 \label{eq:phi-III}
\end{gather}
with
\begin{gather}
  \chi^{\rm I}_{A,B,m}(z) = \chi^{\rm II}_{A,B,m}(z) = (1-z)^{\frac{1}{2}\left(A-m + \frac{B}{A-m}\right)}
  (1+z)^{\frac{1}{2}\left(A-m - \frac{B}{A-m}\right)},  \label{eq:chi-I}
\\
  \chi^{\rm III}_{A,B,m}(z) = (1-z)^{-\frac{1}{2}\left(A+m+1 + \frac{B}{A+m+1}\right)}
  (1+z)^{-\frac{1}{2}\left(A+m+1 - \frac{B}{A+m+1}\right)}, \label{eq:chi-III}
\end{gather}
and corresponding energies
\begin{gather}
  E^{\rm I}_{A,B,m} = E^{\rm II}_{A,B,m} = - (A-m)^2 - \frac{B^2}{(A-m)^2}, \nonumber\\
   E^{\rm III}_{A,B,m} =
  - (A+m+1)^2 - \frac{B^2}{(A+m+1)^2}.  \label{eq:E-I-II-III}
\end{gather}

{\sloppy From each of such factorization functions, we can construct a superpotential $W(x) = - \bigl(\phi(x)\bigr)'$, giving rise to a pair of partner potentials
\begin{gather*}
  V^{(\pm)}(x) = W^2(x) \mp W'(x) + E.
\end{gather*}
The operators
\begin{gather}
  \hat{A}^{\dagger} = - \frac{d}{dx} + W(x), \qquad \hat{A} = \frac{d}{dx} + W(x)  \label{eq:A}
\end{gather}
lead to two factorized Hamiltonians $H^{(+)} = \hat{A}^{\dagger} \hat{A}$ and $H^{(-)} = \hat{A} \hat{A}^{\dagger}$, which can be expressed as
\begin{gather*}
  H^{(\pm)} = - \frac{d^2}{dx^2} + V^{(\pm)}(x) - E
\end{gather*}
and satisfy the intertwining relations $\hat{A} H^{(+)} = H^{(-)} \hat{A}$ and $\hat{A}^{\dagger} H^{(-)} = H^{(+)} \hat{A}^{\dagger}$. The functions $\phi^{\rm I}_{A,B,m}(x)$ and $\phi^{\rm II}_{A,B,m}(x)$ yield two isospectral partners since their inverse is not normalizable, whereas $\phi^{\rm III}_{A,B,m}(x)$ creates a partner $V^{(-)}(x)$ with an additional bound state below the spectrum of $V^{(+)}(x)$, corresponding to its normalizable inverse.

}

To obtain for $V^{(-)}(x)$ some rationally-extended Rosen--Morse II potential with given~$A$ and~$B$, we have to start from a conventional potential with some dif\/ferent $A'$, but the same~$B$. From equations (\ref{eq:phi-I})--(\ref{eq:E-I-II-III}), it is straightforward to get
\begin{gather}
V^{(+)}(x) = V_{A',B}(x), \qquad V^{(-)}(x) = V_{A,B,{\rm ext}}(x) = V_{A,B}(x) + V_{A,B,{\rm rat}}(x),\nonumber \\
V_{A,B,{\rm rat}}(x) = 2\big(1-z^2\big) \left\{2z \frac{\dot{g}^{(A,B)}_m}{g^{(A,B)}_m} - \big(1-z^2\big) \left[\frac{\ddot{g}^{(A,B)}_m}{g^{(A,B)}_m} - \left(\frac{\dot{g}^{(A,B)}_m}
        {g^{(A,B)}_m}\right)^2\right] - m\right\},
 \label{eq:V+-}
\end{gather}
where a dot denotes a derivative with respect to $z$. According to the choice made for the factorization function $\phi(x)$, we may distinguish the three cases
\begin{alignat}{3}
  & ({\rm I}) \  &&  A' = A+1, \quad \phi = \phi^{\rm I}_{A+1,B,m}, \quad g^{(A,B)}_m(z) =
         P^{(\alpha_m, \beta_m)}_m(z),  & \nonumber \\
  &&& \alpha_m = A+1-m + \frac{B}{A+1-m}, \quad \beta_m = A+1-m - \frac{B}{A+1-m},&  \nonumber \\
  &&& m = 1, 2, 3, \ldots, \quad A > m-1, \quad (A+1)(A+1-m) < B < (A+1)^2;& \label{eq:type-I} \\
  & ({\rm II}) \  &&  A' = A+1, \quad \phi = \phi^{\rm II}_{A+1,B,m}, \quad g^{(A,B)}_m(z) =
         P^{(\alpha_m, \beta_m)}_m(z), & \nonumber \\
  &&& \alpha_m = A+1-m + \frac{B}{A+1-m}, \quad \beta_m = A+1-m - \frac{B}{A+1-m},&  \nonumber \\
  &&& m = 1, 2, 3, \ldots, \quad \frac{1}{2}(m-2) < A < m-1, \quad -(A+1)(A+1-m) < B < (A+1)^2; & \nonumber \\
  & ({\rm III}) \  &&  A' = A-1, \quad \phi = \phi^{\rm III}_{A-1,B,m}, \quad g^{(A,B)}_m(z) =
         P^{(-\alpha_{-m-1}, -\beta_{-m-1})}_m(z), & \nonumber \\
  &&& \alpha_{-m-1} = A+m + \frac{B}{A+m}, \quad \beta_{-m-1} = A+m - \frac{B}{A+m}, &  \nonumber \\
  &&& m = 2, 4, 6, \ldots, \quad A>1, \quad 0 < B < (A-1)^2. & \nonumber
\end{alignat}

\subsection[Type-I rationally-extended Rosen-Morse II potentials]{Type-I rationally-extended Rosen--Morse II potentials}\label{section2.2}

In type I case, $V^{(+)}(x)$ and $V^{(-)}(x)$ are isospectral and their common bound-state spectrum is given by
\begin{gather*}
  E^{(+)}_{\nu} = E^{(-)}_{\nu} = - (A+1-\nu)^2 - \frac{B^2}{(A+1-\nu)^2}, \qquad \nu=0, 1, \ldots,
  \nu_{\max}, \\
   A - \sqrt{B} \le \nu_{\max} < A + 1 - \sqrt{B}.
\end{gather*}
The number of bound states $\nu_{\max} + 1$ may range from one to $m$ according to the values taken by $A$ and $B$. For $m=1$, for instance, it is equal to one for all allowed $A$, $B$ values. For $m=2$, it is one or two according to whether $A \le \sqrt{B} < A+1$ or $\sqrt{(A+1)(A-1)} < \sqrt{B} < A$, respectively. For $m=3$, it is one for $A \le \sqrt{B} < A+1$ and becomes two for either $2 < A < 3$ and $A-1 \le \sqrt{B} < A$ or $A \ge 3$ and $\sqrt{(A+1)(A-2)} < \sqrt{B} < A$. Finally, it is as high as three for $2 < A < 3$ and $\sqrt{(A+1)(A-2)} < \sqrt{B} < A-1$. For higher $m$ values, the maximum number $m$ of bound states is achieved for $m-1 < A < (m^2-3m+3)/(m-2)$ and $\sqrt{(A+1)(A+1-m)} < \sqrt{B} < A+2-m$.

From the bound-state wavefunctions
\begin{gather*}
    \psi^{(+)}_{\nu}(x) \propto (1-z)^{\alpha_{\nu}/2} (1+z)^{\beta _{\nu}/2}
       P^{(\alpha_{\nu}, \beta_{\nu})}_{\nu}(z), \qquad \nu=0, 1, \ldots, \nu_{\max}, \\
    \alpha_{\nu} = A+1-\nu + \frac{B}{A+1-\nu}, \qquad \beta_{\nu} = A+1-\nu - \frac{B}{A+1-\nu},
\end{gather*}
of $V^{(+)}(x)$, those of $V^{(-)}(x)$ are obtained by applying the operator $\hat{A}$ given in (\ref{eq:A}), namely
\begin{gather}
  \hat{A}  = \big(1-z^2\big) \frac{d}{dz} + \frac{B}{A+1-m} + (A+1-m) z - \big(1-z^2\big) \frac{\dot{g}^{(A,B)}_m}{g^{(A,B)}_m}
       \nonumber\\
\hphantom{\hat{A}}{}
= \big(1-z^2\big) \frac{d}{dz} + \frac{B}{A+1} + (A+1) z - \frac{2(m+\alpha_m)(m+\beta_m)}
      {2m+\alpha_m+\beta_m} \frac{g^{(A-1,B)}_{m-1}}{g^{(A,B)}_m}.
  \label{eq:A-typeI}
\end{gather}
In going from the f\/irst to the second line of (\ref{eq:A-typeI}), we have used the def\/inition of $g^{(A,B)}_m(z)$, given in (\ref{eq:type-I}), as well as equation (8.961.3) of \cite{gradshteyn}. The results read
\begin{gather}
  \psi^{(-)}_{\nu}(x) \propto \frac{(1-z)^{\alpha_{\nu}/2} (1+z)^{\beta _{\nu}/2}}{g^{(A,B)}_m(z)} y^{(A,B)}_n(z),
  \qquad n=m + \nu - 1, \quad \nu=0, 1, \ldots, \nu_{\max},\!\!\!  \label{eq:psi-typeI}
\end{gather}
where $y^{(A,B)}_n(z)$ is some $n$th-degree polynomial in $z$, def\/ined by
\begin{gather*}
  y^{(A,B)}_n(z)  = \frac{2(\nu+\alpha_{\nu})(\nu+\beta_{\nu})}{2\nu+\alpha_{\nu}+\beta_{\nu}} g^{(A,B)}_m(z)
       P^{(\alpha_{\nu}, \beta_{\nu})}_{\nu-1}(z) \\
\hphantom{y^{(A,B)}_n(z)  =}{}
 - \frac{2(m+\alpha_m)(m+\beta_m)}{2m+\alpha_m+\beta_m}
       g^{(A-1,B)}_{m-1}(z) P^{(\alpha_{\nu}, \beta_{\nu})}_{\nu}(z),
\end{gather*}
where use has been made of the same equation of \cite{gradshteyn}.

As a special case, the ground-state wavefunction of $V^{(-)}(x)$ can be written as
\begin{gather}
  \psi^{(-)}_0(x) \propto \frac{(1-z)^{\alpha_0/2} (1+z)^{\beta _0/2}}{g^{(A,B)}_m(z)} g^{(A-1,B)}_{m-1}(z).
  \label{eq:psi-0-typeI}
\end{gather}
It is worth observing here that from the condition $(A+1) (A+1-m) < B$, responsible for the absence of zeros in $g^{(A,B)}_m(z)$, it follows that $A (A+1-m) < B$, so that $g^{(A-1,B)}_{m-1}(z)$ is also a~nonvanishing polynomial in $(-1, +1)$, as it should be.

On the other hand, by directly inserting (\ref{eq:psi-typeI}) in the Schr\"odinger equation for $V^{(-)}(x)$, we arrive at the following second-order
dif\/ferential equation for $y^{(A,B)}_{m+\nu-1}(z)$,
\begin{gather}
    \left\{\big(1-z^2\big) \frac{d^2}{dz^2} - \left[\alpha_{\nu} - \beta_{\nu} + (\alpha_{\nu} + \beta_{\nu} + 2)z
        + 2\big(1-z^2\big) \frac{\dot{g}^{(A,B)}_m}{g^{(A,B)}_m}\right] \frac{d}{dz}\right. \nonumber \\
    \qquad{} + (\nu-1) (\alpha_{\nu} + \beta_{\nu} + \nu) - m (\alpha_m + \beta_m + m -1) \nonumber \\
   \left. \qquad{} + [\alpha_{\nu} - \beta_{\nu} + \alpha_m - \beta_m + (\alpha_{\nu}
        + \beta_{\nu} + \alpha_m + \beta_m)z] \frac{\dot{g}^{(A,B)}_m}{g^{(A,B)}_m} \right\}
        y^{(A,B)}_{m+\nu-1}(z) = 0, \nonumber \\
    \qquad \nu=0, 1, \ldots, \nu_{\max}. \label{eq:diff-eq}
\end{gather}

As illustrations, let us quote some results obtained for the rational part of the extended potentials
\begin{gather}
  V_{A,B,{\rm rat}}(x) = \frac{N_1(x)}{D(x)} + \frac{N_2(x)}{D^2(x)} + C.  \label{eq:V-rat}
\end{gather}
For $m=1$, we obtain
\begin{gather}
  N_1(x)   = \frac{4B}{A^2 (A+1)^2} \big[A^2 (A+1)^2 - B^2\big], \qquad
  N_2(x)   = \frac{2}{A^2 (A+1)^2} \big[A^2 (A+1)^2 - B^2\big]^2, \nonumber\\
  D(x)   = A(A+1) \tanh x + B, \qquad
  C  = -\frac{2}{A^2 (A+1)^2} \big[A^2 (A+1)^2 - B^2\big],
   \label{eq:V-rat-1}
\end{gather}
with $A > 0$ and $A(A+1) < B < (A+1)^2$, while for $m=2$, we get
\begin{gather}
  N_1(x)  = -16 \frac{B^2 - (A-1)^2 (A+1)^2}{(A-1)^2 (A+1)^3 (2A+1)} \big[(A-1)^2 (A+1) (2A+1) B \tanh x \nonumber\\
\hphantom{N_1(x)  =}{} + \big(A^2+4A+1\big) B^2 + A (A-1)^3 (A+1)^2\big], \nonumber\\
  N_2(x)  = 32 \frac{[B^2 - (A-1)^2 (A+1)^2]^2}{(A-1)^2 (A+1)^3 (2A+1)} \big[2A (A-1) (2A+1) B \tanh x +
      (3A+1) B^2 \nonumber\\
\hphantom{N_2(x)  =}{}
 + A^2 (A-1)^2 (A+1)\big], \nonumber\\
  D(x)  = (A-1)^2 (A+1) (2A+1) \tanh^2 x + 2(A-1) (2A+1) B \tanh x \nonumber\\
\hphantom{D(x)  =}{}  + 2B^2 - (A-1)^2 (A+1), \nonumber\\
  C  = 8A \frac{B^2 - (A-1)^2 (A+1)^2}{(A-1)^2 (A+1)^2 (2A+1)},
   \label{eq:V-rat-2}
\end{gather}
with $A > 1$ and $(A+1) (A-1) < B < (A+1)^2$.

\subsection[Type II rationally-extended Rosen-Morse II potentials]{Type II rationally-extended Rosen--Morse II potentials}\label{section2.3}

The results for type II case only dif\/fering from those for type I in the range of parame\-ters~$A$,~$B$, which is now $\frac{m-2}{2} < A < m-1$ and $- (A+1) (A+1-m) < B < (A+1)^2$, all equations given in Subsection~\ref{section2.2} remain valid. The only change is in the dependence of the number of bound states upon $m$, $A$, and $B$. For $m=1$, it is equal to one for all allowed $A$, $B$ values again. However, for $m=2$, it is one if either $0 < A \le \frac{1}{\sqrt{2}}$ and $\sqrt{(1+A)(1-A)} < \sqrt{B} < A+1$ or $\frac{1}{\sqrt{2}} < A < 1$ and $A \le \sqrt{B} < A+1$, while it amounts to two whenever $\frac{1}{\sqrt{2}} < A < 1$ and $\sqrt{(1+A)(1-A)} < \sqrt{B} < A$. For higher $m$ values, it may range from one to $m$, the maximum number being attained in the case where $\frac{1}{4} \bigl[3(m-2) + \sqrt{m^2+4m-4}\bigr] < A < m-1$ and $\sqrt{(1+A) (m-A-1)} < \sqrt{B} < A-m+2$.

\subsection[Type III rationally-extended Rosen-Morse II potentials]{Type III rationally-extended Rosen--Morse II potentials}\label{section2.4}

In type III case, $V^{(+)}(x)$ and $V^{(-)}(x)$ are not isospectral anymore. Their bound-state spectra are given  instead by
\begin{gather*}
  E^{(+)}_{\nu} = - (A-1-\nu)^2 - \frac{B^2}{(A-1-\nu)^2}, \qquad \nu=0, 1, \ldots, \nu_{\max}, \\
   A-2
  - \sqrt{B} \le \nu_{\max} < A - 1 - \sqrt{B},
\end{gather*}
and
\begin{gather*}
  E^{(-)}_{\nu}  = - (A-1-\nu)^2 - \frac{B^2}{(A-1-\nu)^2}, \qquad \nu=-m-1, 0, 1, \ldots, \nu_{\max}, \\
  A-2 - \sqrt{B} \le \nu_{\max} < A - 1 - \sqrt{B},
\end{gather*}
the ground state of $V^{(-)}(x)$ corresponding to $E^{(-)}_{-m-1} = E^{\rm III}_{A-1,B,m} = - (A+m)^2 - \frac{B^2}{(A+m)^2}$. Observe that here the number of bound states $\nu_{\max}+2$ of $V^{(-)}(x)$ does not depend on $m$ and is entirely determined by $A$ and $B$. For a given $A$ value, it may range from two up to the largest integer contained in $A+1$.

The bound-state wavefunctions of $V^{(-)}(x)$ can be written as
\begin{gather}
  \psi^{(-)}_{\nu}(x)  \propto \frac{(1-z)^{\alpha_{\nu}/2} (1+z)^{\beta _{\nu}/2}}{g^{(A,B)}_m(z)} y^{(A,B)}_n(z),
       \qquad n=m + \nu + 1, \nonumber\\
  \nu  =-m-1, 0, 1, \ldots, \nu_{\max},  \nonumber\\
  \alpha_{\nu}  = A-1-\nu + \frac{B}{A-1-\nu}, \qquad \beta_{\nu} = A-1-\nu - \frac{B}{A-1-\nu},
   \label{eq:psi-typeIII}
\end{gather}
where $y^{(A,B)}_n(z)$ is an $n$th-degree polynomial in $z$. For the ground state, on one hand, we have
\begin{gather*}
  \psi^{(-)}_{-m-1}(x)   \propto \left(\phi^{\rm III}_{A-1,B,m}(x)\right)^{-1} \propto \frac{(1-z)^{\alpha_{-m-1}/2}
       (1+z)^{\beta _{-m-1}/2}}{g^{(A,B)}_m(z)},  \qquad
  y_0^{(A,B)}(z)  =1.
\end{gather*}
For the excited states, on the other hand, we can get (\ref{eq:psi-typeIII}) by starting from $\psi^{(-)}_{\nu}(x) \propto \hat{A} \psi^{(+)}_{\nu}(x)$, $\nu=0, 1, \ldots,\nu_{\max}$, where $\psi^{(+)}_{\nu}(x) \propto (1-z)^{\alpha_{\nu}/2} (1+z)^{\beta_{\nu}/2} P_{\nu}^{(\alpha_{\nu}, \beta_{\nu})}(z)$ and
\begin{gather*}
  \hat{A}  = \big(1-z^2\big) \frac{d}{dz} - \frac{B}{A+m} - (A+m) z - \big(1-z^2\big) \frac{\dot{g}^{(A,B)}_m}{g^{(A,B)}_m}
       \\
\hphantom{\hat{A}}{}
= \big(1-z^2\big) \frac{d}{dz} + \frac{B}{A-1} + (A-1) z + \frac{2(m+1)(m-\alpha_{-m-1}-\beta_{-m-1}+1)}
      {2m-\alpha_{-m-1}-\beta_{-m-1}+2} \frac{g^{(A-1,B)}_{m+1}}{g^{(A,B)}_m}.
\end{gather*}
Observe that one goes from the f\/irst to the second line of this equation by using a combination of equations (8.961.2) and (8.961.3) of~\cite{gradshteyn}. Applying $\hat{A}$ on $\psi^{(+)}_{\nu}(x)$ yields
\begin{gather*}
  y^{(A,B)}_n(z)  = \frac{2(\nu+\alpha_{\nu})(\nu+\beta_{\nu})}{2\nu+\alpha_{\nu}+\beta_{\nu}} g^{(A,B)}_m(z)
       P^{(\alpha_{\nu}, \beta_{\nu})}_{\nu-1}(z) \\
\hphantom{y^{(A,B)}_n(z)  =}{}  + \frac{2(m+1)(m-\alpha_{-m-1}-\beta_{-m-1}+1)}{2m-\alpha_{-m-1}-\beta_{-m-1}+2}
       g^{(A-1,B)}_{m+1}(z) P^{(\alpha_{\nu}, \beta_{\nu})}_{\nu}(z),
\end{gather*}
for $\nu=0, 1, \ldots, \nu_{\max}$.

In particular, the f\/irst excited-state wavefunction of $V^{(-)}(x)$ can be written as
\begin{gather*}
  \psi^{(-)}_0(x) \propto \frac{(1-z)^{\alpha_0/2} (1+z)^{\beta _0/2}}{g^{(A,B)}_m(z)} g^{(A-1,B)}_{m+1}(z).
\end{gather*}
From the general rule for the number of zeros of Jacobi polynomials in $(-1, +1)$, given in (\ref{eq:zeros-1})--(\ref{eq:zeros-3}), it can be easily checked that $g^{(A-1,B)}_{m+1}(z) = P_{m+1}^{(-\alpha_{-m-1}, - \beta_{-m-1})}(z)$ has one zero in this interval, as it should be for the polynomial part of a f\/irst excited-state wavefunction.

Instead of equation (\ref{eq:diff-eq}), $y^{(A,B)}_n(z)$ now fulf\/ils the second-order dif\/ferential equation
\begin{gather}
   \left\{\big(1-z^2\big) \frac{d^2}{dz^2} - \left[\alpha_{\nu} - \beta_{\nu} + (\alpha_{\nu} + \beta_{\nu} + 2)z
        + 2\big(1-z^2\big) \frac{\dot{g}^{(A,B)}_m}{g^{(A,B)}_m}\right] \frac{d}{dz} \right.\nonumber \\
  \quad {}+ (\nu+1) (\alpha_{\nu} + \beta_{\nu} + \nu + 2) - m (m - \alpha_{-m-1} - \beta_{-m-1} -1) \nonumber \\
 \left. \quad {} + [\alpha_{\nu} - \beta_{\nu} - \alpha_{-m-1} + \beta_{-m-1} + (\alpha_{\nu}
        + \beta_{\nu} - \alpha_{-m-1} - \beta_{-m-1})z] \frac{\dot{g}^{(A,B)}_m}{g^{(A,B)}_m}\! \right\}\!
        y^{(A,B)}_{m+\nu+1}(z) = 0, \nonumber \\
  \quad \nu=-m-1, 0, 1, \ldots, \nu_{\max}. \label{eq:diff-eq-bis}
\end{gather}

The results obtained here may be illustrated by considering the lowest allowed $m$ value, namely $m=2$. In such a case, the rational part of the extended potential takes the form (\ref{eq:V-rat}) with
\begin{gather}
  N_1(x)  = -16 \frac{B^2 - A^2 (A+2)^2}{A^3 (A+2)^2 (2A+1)} \big[A (A+2)^2 (2A+1) B \tanh x \nonumber\\
\hphantom{N_1(x)  =}{} + \big(A^2-2A-2\big) B^2 + A^2 (A+1) (A+2)^3\big], \nonumber\\
  N_2(x)  = -32 \frac{[B^2 - A^2 (A+2)^2]^2}{A^3 (A+2)^2 (2A+1)} \big[2(A+1) (A+2) (2A+1) B \tanh x +
      (3A+2) B^2 \nonumber\\
 \hphantom{N_2(x)  =}{}
  + A (A+1)^2 (A+2)^2\big], \nonumber\\
  D(x)  = A (A+2)^2 (2A+1) \tanh^2 x + 2(A+2) (2A+1) B \tanh x  + 2B^2 + A (A+2)^2, \nonumber\\
  C  = 8(A+1) \frac{B^2 - A^2 (A+2)^2}{A^2 (A+2)^2 (2A+1)},
   \label{eq:V-rat-2-bis}
\end{gather}
where $A > 1$ and $0 < B < (A-1)^2$.

\section{Rationally-extended Eckart potentials\\ in f\/irst-order SUSYQM}\label{section3}

\subsection{General results}\label{section3.1}

The Eckart potential
\begin{gather}
  V_{A,B}(x) = A(A-1) \operatorname{csch}^2 x - 2B \coth x, \qquad 0 < x < \infty,  \label{eq:Eckart}
\end{gather}
where we assume $A > 1$ and $B > A^2$,\footnote{Note that the assumption $A>1$ ensures that the potential is repulsive for $x \to 0$ while the hypothesis $B>A^2$ is necessary for getting at least one bound state.} has a f\/inite number of bound states, whose energy and wavefunction can be expressed as (see, e.g.,~\cite{cooper})
\begin{gather*}
  E^{(A,B)}_{\nu} = - (A + \nu)^2 - \frac{B^2}{(A + \nu)^2}, \qquad \nu = 0, 1, \ldots, \nu_{\max}, \\
  \sqrt{B} - A - 1 \le \nu_{\max} < \sqrt{B} - A,
\end{gather*}
and
\begin{gather*}
  \psi^{(A,B)}_{\nu}(x)  \propto (\sinh x)^{A+\nu} \exp \left(- \frac{B}{A+\nu} x\right)
      P_{\nu}^{\left(-A-\nu+\frac{B}{A+\nu}, -A-\nu-\frac{B}{A+\nu}\right)}(\coth x) \\
\hphantom{\psi^{(A,B)}_{\nu}(x)}{}
\propto (z-1)^{-\frac{1}{2}\left(A+\nu-\frac{B}{A+\nu}\right)} (z+1)^{-\frac{1}{2}\left(A+\nu+\frac{B}{A+\nu}
      \right)} P_{\nu}^{\left(-A-\nu+\frac{B}{A+\nu}, -A-\nu-\frac{B}{A+\nu}\right)}(z),
\end{gather*}
respectively. Here $z = \coth x$ varies in the interval $1 < z < \infty$.

In terms of such a variable, the corresponding Schr\"odinger equation can be written as
\begin{gather}
  \left[- \big(1-z^2\big)^2 \frac{d^2}{dz^2} + 2z \big(1-z^2\big) \frac{d}{dz} + A(A-1) \big(z^2-1\big) - 2B z\right] \phi\bigl(x(z)\bigr)
  = E \phi\bigl(x(z)\bigr).  \label{eq:SE-ter}
\end{gather}
Formally, it can be obtained from (\ref{eq:SE-bis}), valid for the Rosen--Morse II potential, by substituting $(-A, -B)$ for $(A,B)$. Hence, we can directly infer that equation (\ref{eq:SE-ter}) admits the two independent polynomial-type solutions
\begin{gather}
  \phi_1(x)  = (z-1)^{-\frac{1}{2}\left(A+m - \frac{B}{A+m}\right)} (z+1)^{-\frac{1}{2}\left(A+m + \frac{B}{A+m}
      \right)} P_m^{\left(-A-m + \frac{B}{A+m}, -A-m - \frac{B}{A+m}\right)}(z), \nonumber\\
  E_1  = - (A+m)^2 - \frac{B^2}{(A+m)^2},
  \label{eq:phi-1-bis}
\end{gather}
and
\begin{gather}
  \phi_2(x)  = (z-1)^{\frac{1}{2}\left(A-m-1 - \frac{B}{A-m-1}\right)} (z+1)^{\frac{1}{2}\left(A-m-1
      + \frac{B}{A-m-1}\right)} \nonumber\\
 \hphantom{\phi_2(x)  =}{}
    \times  P_m^{\left(A-m-1 - \frac{B}{A-m-1}, A-m-1 + \frac{B}{A-m-1}\right)}(z), \nonumber\\
  E_2  = - (A-m-1)^2 - \frac{B^2}{(A-m-1)^2}.
  \label{eq:phi-2-bis}
\end{gather}

The conditions on $A$ and $B$ such that these solutions correspond to an energy eigenvalue below the ground-state one, $E_0^{(A,B)} = - A^2 - \frac{B^2}{A^2}$, dif\/fer, however, from those found in Section~\ref{section2} due to some changes in their admissible values. For the f\/irst solution (\ref{eq:phi-1-bis}), we f\/ind a single possibility
\begin{alignat*}{3}
  & ({\rm 1a}) \quad && A > 1, \quad A^2 < B < A(A+m).&
\end{alignat*}
In contrast, for the second solution (\ref{eq:phi-2-bis}), we get the three cases
\begin{alignat*}{3}
  & ({\rm 2a}) \quad & &  A > m+1, \quad B > A^2; & \\
  & ({\rm 2b}) \quad & & \tfrac{m+1}{2} < A < m+1, \qquad B > A^2; & \\
  & ({\rm 2c}) \quad &&  1 < A < \tfrac{m+1}{2}, \qquad A^2 < B < - A(A-m-1). &
\end{alignat*}

Other discrepancies with respect to the Rosen--Morse II potential arise when checking the absence of zeros of the Jacobi polynomials in (\ref{eq:phi-1-bis}) and (\ref{eq:phi-2-bis}). Since their variable~$z$ now varies in the interval $(1, \infty)$, we have to use Rule~2 of Appendix~\ref{appendixA} instead of Rule~1. For the f\/irst solution~$\phi_1(x)$, on one hand, it can be easily shown that the parameters $\alpha$, $\beta$ in~(\ref{eq:phi-1-bis}) are such that $\alpha < -m$, $\alpha + \beta < -2m$, which corresponds to Case~a of Rule~2. For the second solution~$\phi_2(x)$, on the other hand, the parameters $\alpha$, $\beta$ in (\ref{eq:phi-2-bis}) satisfy the conditions~$\alpha < -m$, $\alpha + \beta > 0$ in Case~2a, $\alpha > 0$, $\alpha + \beta > -m-1$ in Case~2b, and $-m < \alpha < 0$, $-2m < \alpha + \beta < -m-1$ in Case~2c. It is therefore clear that the second possibility is associated with Case~b of Rule~2, while the f\/irst one corresponds to Case~c (nonexceptional subcase) of the same provided~$m$ is chosen even ($m=2k$). Finally, the third occurrence may agree with some exceptional subcases of Case~c for appropriately chosen parameters.

We conclude that, apart from some exceptional cases, which will not be considered any further, there exist three acceptable polynomial-type, nodeless solutions of the Eckart potential Schr\"odinger equation,
\begin{gather*}
  \phi^{\rm I}_{A,B,m}(x)  = \chi^{\rm I}_{A,B,m}(z) P_m^{\left(-A-m + \frac{B}{A+m}, -A-m - \frac{B}{A+m}\right)}
       (z) \\
\hphantom{\phi^{\rm I}_{A,B,m}(x)  =}{} \ \text{if} \ m=1, 2, 3, \ldots, \quad A > 1, \quad A^2 < B < A(A+m),
\\
  \phi^{\rm II}_{A,B,m}(x)  = \chi^{\rm II}_{A,B,m}(z) P_m^{\left(A-m-1 - \frac{B}{A-m-1}, A-m-1 + \frac{B}{A-m-1}
       \right)}(z) \\
\hphantom{\phi^{\rm II}_{A,B,m}(x)  =}{} \  \text{if} \ m=1, 2, 3, \ldots, \quad \tfrac{m+1}{2} < A < m+1, \quad B > A^2,
\\
  \phi^{\rm III}_{A,B,m}(x)  = \chi^{\rm III}_{A,B,m}(z)
       P_m^{\left(A-m-1 - \frac{B}{A-m-1}, A-m-1 + \frac{B}{A-m-1}\right)}(z) \\
\hphantom{}{} \ \text{if} \ m=2, 4, 6, \ldots, \quad A > m+1, \quad B > A^2,
\end{gather*}
with
\begin{gather*}
  \chi^{\rm I}_{A,B,m}(z) = (z-1)^{-\frac{1}{2}\left(A+m - \frac{B}{A+m}\right)}
  (z+1)^{-\frac{1}{2}\left(A+m + \frac{B}{A+m}\right)},
\\
  \chi^{\rm II}_{A,B,m}(z) = \chi^{\rm III}_{A,B,m}(z) = (z-1)^{\frac{1}{2}\left(A-m-1 - \frac{B}{A-m-1}\right)}
  (z+1)^{\frac{1}{2}\left(A-m-1 + \frac{B}{A-m-1}\right)},
\end{gather*}
and corresponding energies
\begin{gather*}
  E^{\rm I}_{A,B,m} = - (A+m)^2 - \frac{B^2}{(A+m)^2}, \\
   E^{\rm II}_{A,B,m} = E^{\rm III}_{A,B,m} =
  - (A-m-1)^2 - \frac{B^2}{(A-m-1)^2}.
\end{gather*}
In contrast with the f\/irst two solutions, the third one has a normalizable inverse on the positive half-line.

On considering such factorization functions in f\/irst-order SUSYQM and proceeding as in Section~\ref{section2}, we arrive at a pair of partner potentials, which are still given by equation~(\ref{eq:V+-}) with~$V_{A,B}(x)$ denoting the Eckart potential~(\ref{eq:Eckart}) and any one of the following three possibilities
\begin{alignat}{3}
  & ({\rm I}) \ & &  A' = A-1, \quad \phi = \phi^{\rm I}_{A-1,B,m}, \quad g^{(A,B)}_m(z) =
         P^{(\alpha_m, \beta_m)}_m(z),  & \nonumber \\
  &&&  \alpha_m = -A+1-m + \frac{B}{A-1+m}, \quad \beta_m = -A+1-m - \frac{B}{A-1+m}, & \nonumber \\
  &&&  m = 1, 2, 3, \ldots, \quad A > 2, \quad (A-1)^2 < B < (A-1)(A-1+m); & \label{eq:type-I-bis} \\
  & ({\rm II}) \ &&  A' = A+1, \quad \phi = \phi^{\rm II}_{A+1,B,m}, \quad g^{(A,B)}_m(z) =
         P^{(-\alpha_{-m-1}, -\beta_{-m-1})}_m(z),  & \nonumber \\
  &&&  \alpha_{-m-1} = -A+m + \frac{B}{A-m}, \quad \beta_{-m-1} = -A+m - \frac{B}{A-m}, & \nonumber \\
  &&& m = 1, 2, 3, \ldots, \quad \frac{m-1}{2} < A < m, \quad B > (A+1)^2; & \label{eq:type-II-bis} \\
  & ({\rm III}) \ \ &&  A' = A+1, \quad \phi = \phi^{\rm III}_{A+1,B,m}, \quad g^{(A,B)}_m(z) =
         P^{(-\alpha_{-m-1}, -\beta_{-m-1})}_m(z), & \nonumber \\
  &&& \alpha_{-m-1} = -A+m + \frac{B}{A-m}, \quad \beta_{-m-1} = -A+m - \frac{B}{A-m}, & \nonumber \\
  &&& m = 2, 4, 6, \ldots, \quad A>m, \quad B > (A+1)^2. & \nonumber
\end{alignat}

\subsection{Type-I rationally-extended Eckart potentials}\label{section3.2}

The partner potentials $V^{(\pm)}(x)$ are isospectral with a bound-state spectrum given by
\begin{gather*}
  E^{(+)}_{\nu} = E^{(-)}_{\nu} = - (A-1+\nu)^2 - \frac{B^2}{(A-1+\nu)^2}, \qquad \nu=0, 1, \ldots,
  \nu_{\max}, \\
   \sqrt{B} - A \le \nu_{\max} < \sqrt{B} - A + 1.
\end{gather*}
The number of bound states $\nu_{\max} + 1$ may range from one to $\left[\frac{m+1}{2}\right]$ according to the values taken by $A$ and $B$. For $m=1$ or~2, for instance, it is equal to one for all allowed $A$, $B$ values. For $m=3$ or~4, it is one for $A-1 < \sqrt{B} \le A$, but two for $A < \sqrt{B} < \sqrt{(A-1)(A+2)}$ or $A < \sqrt{B} < \sqrt{(A-1)(A+3)}$, respectively. For higher $m$ values equal to $2k-1$ or $2k$, the maximum number $k$ of bound states is achieved if $A > k^2 - 2k + 2$ and $A+k-2 < \sqrt{B} < \sqrt{(A-1)(A+2k-2)}$ or if $A > \frac{1}{2}(k^2 - 2k + 3)$ and $A+k-2 < \sqrt{B} < \sqrt{(A-1)(A+2k-1)}$.

On acting with the operator
\begin{gather*}
  \hat{A} = \big(1-z^2\big) \frac{d}{dz} + \frac{B}{A-1+m} - (A-1+m) z - \big(1-z^2\big) \frac{\dot{g}^{(A,B)}_m}{g^{(A,B)}_m}
       \\
\hphantom{\hat{A}}{}
= \big(1-z^2\big) \frac{d}{dz} + \frac{B}{A-1} - (A-1) z - \frac{2(m+\alpha_m)(m+\beta_m)}
      {2m+\alpha_m+\beta_m} \frac{g^{(A+1,B)}_{m-1}}{g^{(A,B)}_m},
\end{gather*}
on the bound-state wavefunctions of $V^{(+)}(x)$, we get for those of $V^{(-)}(x)$
\begin{gather*}
  \psi^{(-)}_{\nu}(x)  \propto \frac{(z-1)^{\alpha_{\nu}/2} (z+1)^{\beta _{\nu}/2}}{g^{(A,B)}_m(z)} y^{(A,B)}_n(z),
\qquad n=m + \nu - 1, \qquad \nu=0, 1, \ldots, \nu_{\max}, \\
  \alpha_{\nu}  = -A+1-\nu + \frac{B}{A-1+\nu}, \qquad \beta_{\nu} = -A+1-\nu - \frac{B}{A-1+\nu}.
\end{gather*}
Here $y^{(A,B)}_n(z)$ is some $n$th-degree polynomial in $z$, def\/ined by
\begin{gather*}
  y^{(A,B)}_n(z)  = \frac{2(\nu+\alpha_{\nu})(\nu+\beta_{\nu})}{2\nu+\alpha_{\nu}+\beta_{\nu}} g^{(A,B)}_m(z)
       P^{(\alpha_{\nu}, \beta_{\nu})}_{\nu-1}(z) \\
\hphantom{y^{(A,B)}_n(z)  =}{} - \frac{2(m+\alpha_m)(m+\beta_m)}{2m+\alpha_m+\beta_m}
       g^{(A+1,B)}_{m-1}(z) P^{(\alpha_{\nu}, \beta_{\nu})}_{\nu}(z)
\end{gather*}
and satisfying a second-order dif\/ferential equation similar to (\ref{eq:diff-eq}).

In particular, the ground-state wavefunction of $V^{(-)}(x)$ can be expressed as
\begin{gather}
  \psi^{(-)}_0(x) \propto \frac{(z-1)^{\alpha_0/2} (z+1)^{\beta _0/2}}{g^{(A,B)}_m(z)} g^{(A+1,B)}_{m-1}(z),
  \label{eq:psi-0-typeI-bis}
\end{gather}
where the polynomial $g^{(A+1,B)}_{m-1}(z)$ has no zero in $(1, \infty)$, as it should be, because the condition $B < A(A-1+m)$ necessary for ensuring this property is implied by the corresponding one for $g^{(A,B)}_m(z)$, namely $B < (A-1)(A-1+m)$.

The rational part of the extended potentials still takes a form similar to equation~(\ref{eq:V-rat}), where, for $m=1$ and $m=2$, $N_1(x)$, $N_2(x)$, $D(x)$, and $C$ can be inferred from equations~(\ref{eq:V-rat-1}) and~(\ref{eq:V-rat-2}) after substituting $(-A, -B, \coth x)$ for $(A, B, \tanh x)$. The resulting expressions are valid for $A>2$ and $(A-1)^2 < B < (A-1)A$ or $(A-1)^2 < B < (A-1)(A+1)$, respectively.

\subsection{Type II rationally-extended Eckart potentials}\label{section3.3}

In contrast with what happens for the Rosen--Morse II potential, where types~I and~II only dif\/fer in the range of parameters $A$ and $B$, there is here a drastic change in going from type~I to type~II (see equations~(\ref{eq:type-I-bis}) and~(\ref{eq:type-II-bis})). Although the partner potentials $V^{(\pm)}(x)$ remain isospectral with common spectrum given by
\begin{gather}
   E^{(+)}_{\nu} = E^{(-)}_{\nu} = - (A+1+\nu)^2 - \frac{B^2}{(A+1+\nu)^2}, \qquad \nu=0, 1, \ldots,
       \nu_{\max}, \nonumber\\
 \sqrt{B} - A - 2 \le \nu_{\max} < \sqrt{B} - A - 1,  \label{eq:spectrum}
\end{gather}
the number of bound states $\nu_{\max}+1$ is now entirely determined by the $B$ value for a given $A$, independently of $m$. Hence, even for $m=1$, it may be arbitrarily large.

From the operator
\begin{gather*}
  \hat{A}  = (1-z^2) \frac{d}{dz} - \frac{B}{A-m} + (A-m) z - (1-z^2) \frac{\dot{g}^{(A,B)}_m}{g^{(A,B)}_m}
       \\
  \hphantom{\hat{A}}{}
  = (1-z^2) \frac{d}{dz} + \frac{B}{A+1} - (A+1) z + \frac{2(m+1)(m-\alpha_{-m-1}-\beta_{-m-1}+1)}
      {2m-\alpha_{-m-1}-\beta_{-m-1}+2} \frac{g^{(A+1,B)}_{m+1}}{g^{(A,B)}_m},
\end{gather*}
the partner bound-state wavefunctions are obtained in the form
\begin{gather}
  \psi^{(-)}_{\nu}(x)  \propto \frac{(z-1)^{\alpha_{\nu}/2} (z+1)^{\beta _{\nu}/2}}{g^{(A,B)}_m(z)} y^{(A,B)}_n(z),
       \qquad n=m + \nu + 1, \qquad \nu=0, 1, \ldots, \nu_{\max}, \nonumber \\
  \alpha_{\nu}  = -A-1-\nu + \frac{B}{A+1+\nu}, \qquad \beta_{\nu} = -A-1-\nu - \frac{B}{A+1+\nu},
 \label{eq:psi-typeII-bis}
\end{gather}
with
\begin{gather}
  y^{(A,B)}_n(z)  = \frac{2(\nu+\alpha_{\nu})(\nu+\beta_{\nu})}{2\nu+\alpha_{\nu}+\beta_{\nu}} g^{(A,B)}_m(z)
       P^{(\alpha_{\nu}, \beta_{\nu})}_{\nu-1}(z) \nonumber\\
\hphantom{y^{(A,B)}_n(z)  =}{}
 + \frac{2(m+1)(m-\alpha_{-m-1}-\beta_{-m-1}+1)}{2m-\alpha_{-m-1}-\beta_{-m-1}+2}
       g^{(A+1,B)}_{m+1}(z) P^{(\alpha_{\nu}, \beta_{\nu})}_{\nu}(z),
 \label{eq:y-typeII-bis}
\end{gather}
satisfying a dif\/ferential equation similar to~(\ref{eq:diff-eq-bis}). It is worth stressing that the degree of the polynomial $y^{(A,B)}_n(z)$ is now $n=m+\nu+1$, instead of $n=m+\nu-1$ for type~I.

In particular, for the ground-state wavefunction, we get
\begin{gather}
  \psi^{(-)}_0(x) \propto \frac{(z-1)^{\alpha_0/2} (z+1)^{\beta _0/2}}{g^{(A,B)}_m(z)} g^{(A+1,B)}_{m+1}(z),
  \label{eq:psi-0-typeII-bis}
\end{gather}
where $g^{(A+1,B)}_{m+1}(z)$ has no zero in $(1, \infty)$ because the condition $A > \frac{1}{2}(m-2)$ ensuring such a~property is implied by the inequality $A > \frac{1}{2}(m-1)$ given in~(\ref{eq:type-II-bis}).

\subsection{Type III rationally-extended Eckart potentials}\label{section3.4}

The results for type III dif\/fer from those for type II in the range of parameter $A$, which is now $A>m$, and in the restriction of $m$ to even values.

The only important change with respect to Subsection~\ref{section3.3} is that in the present case $V^{(-)}(x)$ exhibits an extra bound state below the spectrum of $V^{(+)}(x)$, associated with the normalizable inverse of the factorization function $\phi^{\rm III}_{A+1,B,m}(x)$. Hence the common part of the $V^{(\pm)}(x)$ spectrum is still given by equation (\ref{eq:spectrum}), but for $V^{(-)}(x)$, the $\nu$ index may also take the value $-m-1$, giving rise to the ground-state energy and wavefunction
\begin{gather*}
  E^{(-)}_{-m-1} = E^{\rm III}_{A+1,B,m} = - (A-m)^2 - \frac{B^2}{(A-m)^2}
\end{gather*}
and
\begin{gather*}
  \psi^{(-)}_{-m-1}(x) \propto \left(\phi^{\rm III}_{A+1,B,m}(x)\right)^{-1} = \frac{(z-1)^{\alpha_{-m-1}}
  (z+1)^{\beta_{-m-1}}}{g^{(A,B)}_m(z)},
\end{gather*}
corresponding to $y^{(A,B)}_0(z) = 1$.

Equations (\ref{eq:psi-typeII-bis}) and (\ref{eq:y-typeII-bis}) now provide us with the excited-state wavefunctions of $V^{(-)}(x)$. As a consequence, equation~(\ref{eq:psi-0-typeII-bis}) describes the f\/irst-excited wavefunction. In accordance with such a property, it can be readily checked from equations~(\ref{eq:zeros-3}), (\ref{eq:zeros-5}), and~(\ref{eq:zeros-6}) that its polynomial part $g^{(A+1,B)}_{m+1}(x) = P^{(-\alpha_{-m-1}, -\beta_{-m-1})}_{m+1}(z)$ has one zero in the interval  $(1, \infty)$ for the choice of parameters pertinent to type III potentials.

Equations (\ref{eq:V-rat}) and (\ref{eq:V-rat-2-bis}) yield an example of type~III potential corresponding to $m=2$ if we replace $A$, $B$, and $\tanh x$ by $-A$, $-B$, and $\coth x$, respectively, and restrict ourselves to $A>2$ and $B>(A+1)^2$.

\section[Enlarged shape invariance property of extended Rosen-Morse II and Eckart potentials]{Enlarged shape invariance property\\ of extended Rosen--Morse II and Eckart potentials}\label{section4}

The purpose of the present Section is to determine the partner $\bar{V}^{(-)}(x)$ of the rationally-extended Rosen--Morse II and Eckart potentials of type~I or~II, $\bar{V}^{(+)}(x) = V^{(-)}(x) = V^{(m)}_{A,B,\rm{ext}}(x)$, when the ground state of the latter is deleted. Here we have appended a superscript $(m)$ to specify the degree of the polynomial $g^{(A,B)}_m(z)$ arising in the denominator of equation~(\ref{eq:V+-}).

In this process, the new superpotential $\bar{W}(x) = - \bigl(\log \psi^{(-)}_0(x)\bigr)'$, obtained from equations (\ref{eq:psi-0-typeI}), (\ref{eq:psi-0-typeI-bis}), or (\ref{eq:psi-0-typeII-bis}), is given by
\begin{gather*}
  \bar{W}(x) = \begin{cases}
    \dfrac{B}{A+1} + (A+1)z - (1-z^2) \left(\dfrac{\dot{g}^{(A-1,B)}_{m-1}}{g^{(A-1,B)}_{m-1}}
      - \dfrac{\dot{g}^{(A,B)}_m}{g^{(A,B)}_m}\right) & \begin{array}{@{}l}
      \text{for Rosen--Morse II} \\ \text{(type I or II)},\end{array} \vspace{1mm}\\
    \dfrac{B}{A-1} - (A-1)z - (1-z^2) \left(\dfrac{\dot{g}^{(A+1,B)}_{m-1}}{g^{(A+1,B)}_{m-1}}
      - \dfrac{\dot{g}^{(A,B)}_m}{g^{(A,B)}_m}\right) & \text{for Eckart (type I)}, \vspace{1mm}\\
    \dfrac{B}{A+1} - (A+1)z - (1-z^2) \left(\dfrac{\dot{g}^{(A+1,B)}_{m+1}}{g^{(A+1,B)}_{m+1}}
      - \dfrac{\dot{g}^{(A,B)}_m}{g^{(A,B)}_m}\right) & \text{for Eckart (type II)}.
  \end{cases}
\end{gather*}
It can be readily seen that the partner $\bar{V}^{(-)}(x) = \bar{V}^{(+)}(x) + 2 \bar{W}'(x)$ is an extended potential of the same type as $\bar{V}^{(+)}(x)$, but with a dif\/ferent parameter $A$ and a dif\/ferent polynomial degree~$m$,
\begin{gather*}
  \bar{V}^{(-)}(x) = \begin{cases}
    V^{(m-1)}_{A-1,B,\rm{ext}}(x) & \text{for Rosen--Morse II (type I or II)}, \\
    V^{(m-1)}_{A+1,B,\rm{ext}}(x) & \text{for Eckart (type I)}, \\
    V^{(m+1)}_{A+1,B,\rm{ext}}(x) & \text{for Eckart (type II)}.
  \end{cases}
\end{gather*}
The change in the parameter $A$ is similar to that observed for the corresponding conventional potential, which is known to be translationally SI \cite{cooper}, but the modif\/ication in the degree $m$ points to the existence of an enlarged SI property, valid for some rational extensions.

Furthermore, it is worth observing that the f\/irst step from $V^{(+)}(x)$ to $V^{(-)}(x)$ (using broken SUSYQM) and the second one from $\bar{V}^{(+)}(x) = V^{(-)}(x)$ to $\bar{V}^{(-)}(x)$ (employing unbroken SUSYQM) can be put together \cite{bagchi, andrianov, bagrov, aoyama, fernandez} to arrive at a reducible second-order SUSYQM transformation from a conventional potential to an extended one, $V^{(m-1)}_{A-1,B,\rm{ext}}(x)$, $V^{(m-1)}_{A+1,B,\rm{ext}}(x)$, or $V^{(m+1)}_{A+1,B,\rm{ext}}(x)$. In each case, the same result can be obtained along another path by combining the usual unbroken SUSYQM transformation relating two conventional potentials with translated parameter $A$ with a broken one connecting conventional with extended potentials. In the Rosen--Morse~II case (type~I or~II), we get in this way the commutative diagram
\begin{gather*}
  \begin{CD}
  V_{A+1,B}(x) @>\text{unbroken}>> V_{A,B}(x)\\
  @V\text{broken}VV @VV\text{broken}V\\
  V_{A,B,{\rm ext}}^{(m)}(x) @>>\text{unbroken}> V_{A-1,B,{\rm ext}}^{(m-1)}(x)
\end{CD}.
\end{gather*}
Similarly, in the Eckart case, we obtain
\begin{gather*}
  \begin{CD}
  V_{A-1,B}(x) @>\text{unbroken}>> V_{A,B}(x)\\
  @V\text{broken}VV @VV\text{broken}V\\
  V_{A,B,{\rm ext}}^{(m)}(x) @>>\text{unbroken}> V_{A+1,B,{\rm ext}}^{(m-1)}(x)
\end{CD}
\end{gather*}
for type I and
\begin{gather*}
  \begin{CD}
  V_{A+1,B}(x) @>\text{unbroken}>> V_{A+2,B}(x)\\
  @V\text{broken}VV @VV\text{broken}V\\
  V_{A,B,{\rm ext}}^{(m)}(x) @>>\text{unbroken}> V_{A+1,B,{\rm ext}}^{(m+1)}(x)
\end{CD}
\end{gather*}
for type II.

\section{Conclusion}\label{section5}

In the present paper, we have derived all rational extensions of the Rosen--Morse II and Eckart potentials that can be obtained in f\/irst-order SUSYQM by starting from polynomial-type, nodeless solutions of the conventional potential Schr\"odinger equation with an energy below the ground state. These extensions belong to three dif\/ferent types, the f\/irst two being strictly isospectral to a conventional potential with dif\/ferent parameters and the third one having an extra bound state below the spectrum of the latter.

In addition, we have found new examples of the novel enlarged SI property, f\/irst pointed out for rational extensions of the Morse potential \cite{cq12}. We have indeed proved that the partner of rationally-extended Rosen--Morse II and Eckart potentials of type I or II, resulting from the deletion of their ground state, can be obtained by translating both the parameter $A$ (as conventional potentials) and the degree $m$ arising in the denominator. Hence it belongs to the same family of rational extensions, which turns out to be closed.

Whether the enlarged SI, exhibited by some rational extensions of the Morse, Rosen--Morse II, and Eckart potentials, would imply in general exact solvability as does the ordinary SI is still unknown, but would be a very interesting topic for future investivation.

As a f\/inal point, it is worth observing that type III rationally-extended Morse, Rosen--Morse II, and Eckart potentials could also be derived in higher-order SUSYQM by using the Krein--Adler's modif\/ication \cite{krein, adler} of Crum's theorem \cite{crum}, as already done in a similar context elsewhere (see, e.g., \cite{odake11}).

\appendix

\section{Zeros of the general Jacobi polynomials on the real line}\label{appendixA}

Let $P_n^{(\alpha, \beta)}(x)$ denote a general Jacobi polynomial with $n \ge 1$ and $\alpha$, $\beta$ any real numbers with the exceptions of $\alpha = -1, -2, \ldots,-n$, $\beta = -1, -2, \ldots,-n$, and $\alpha + \beta = -n-1, -n-2, \ldots,-2n$. The number of its zeros in $-1 < x < +1$ is given by \cite{szego}
\begin{gather}
  N_1(\alpha, \beta) = \begin{cases}
    2 \left[\dfrac{X+1}{2}\right] &\text{if $\displaystyle (-1)^n \binom{n+\alpha}{n} \binom{n+\beta}{n} > 0$}, \vspace{1mm}\\
    2 \left[\dfrac{X}{2}\right] + 1 &\text{if $\displaystyle (-1)^n \binom{n+\alpha}{n} \binom{n+\beta}{n} < 0$},
  \end{cases}   \label{eq:zeros-1}
\end{gather}
where $X$ is def\/ined by
\begin{gather}
  X = X(\alpha, \beta) = E \bigl[\tfrac{1}{2} (|2n+\alpha+\beta+1| - |\alpha| - |\beta| + 1)\bigr],  \label{eq:zeros-2}
\end{gather}
with
\begin{gather}
  E(u) = \begin{cases}
    0 &\text{if $u \le 0$}, \\
    [u] &\text{if $u > 0$ and $u \ne 1, 2, 3, \ldots$}, \\
    u-1 &\text{if $u=1, 2, 3, \ldots$}.
  \end{cases}  \label{eq:zeros-3}
\end{gather}
From this result, it follows that the necessary and suf\/f\/icient conditions for having no zero in $(-1, +1)$ are
\begin{gather}
  X = 0 \qquad \text{and} \qquad (-1)^n \binom{n+\alpha}{n} \binom{n+\beta}{n} > 0.  \label{eq:conditions}
\end{gather}

On taking into account that $\binom{n+\alpha}{n} > 0$ in any one of the cases
\begin{itemize}\itemsep=0pt
  \item [a)] $\alpha \ge 0$,
  \item [b)] $n = 2k$ and $\alpha < -2k$ or $-2k+2l+1 < \alpha < -2k+2l+2$ for $l=0 , 1, \ldots$, or $k-1$,
  \item [c)] $n = 2k+1$ and $-2k+2l-1 < \alpha < -2k+2l$ for $l=0, 1, \ldots$, or $k$,
\end{itemize}
while $\binom{n+\alpha}{n} < 0$ in the remaining cases, namely
\begin{itemize}\itemsep=0pt
  \item [a)] $n = 2k$ and $-2k+2l < \alpha < -2k+2l+1$ for $l=0, 1, \ldots$, or $k-1$,
  \item [b)] $n = 2k+1$ and $\alpha < -2k-1$ or $-2k+2l < \alpha < -2k+2l+1$ for $l=0, 1, \ldots$, or $k-1$,
\end{itemize}
it is possible to reformulate conditions (\ref{eq:conditions}) in the following convenient way:

\medskip\noindent {\bf Rule 1.} $P_n^{(\alpha,\beta)}(x)$ has no zero in $(-1, +1)$ if and only if $\alpha+\beta \ne -n-1, -n-2, \ldots, -2n$ and one of the following cases occurs:
\begin{itemize}\itemsep=0pt
  \item [a)] $\alpha \ge 0$ and $\beta < -n$,
  \item [b)] $\alpha < -n$ and $\beta \ge 0$,
  \item [c)]
    \begin{itemize}\itemsep=0pt
      \item [(i)] for $n=2k$:
        \begin{itemize}\itemsep=0pt
          \item $\alpha < -2k$ and $\beta < -2k$,
          \item or else $\alpha < -2k$ and $-2l-1 < \beta < -2l$ for $l=0, 1, \ldots$, or $k-1$,
          \item or else $\beta < -2k$ and $-2l-1 < \alpha < -2l$ for $l=0, 1, \ldots$, or $k-1$,
          \item or else $-2l-3 < \alpha < -2l-2$ for $l=0$, 1, \ldots, or $k-2$ and $-2m-3 < \beta < -2m-2$ for
            $m=k-l-2, k-l-1, \ldots$, or $k-2$,
          \item or else $-2l-2 < \alpha < -2l-1$ for $l=0, 1, \ldots$, or $k-1$ and $-2m-2 < \beta < -2m-1$ for
            $m=k-l-1, k-l, \ldots$, or $k-1$,
        \end{itemize}
      \item[(ii)] for $n=2k+1$:
        \begin{itemize}\itemsep=0pt
          \item $\alpha < -2k-1$ and $-2l-1 < \beta < -2l$ for $l=0, 1, \ldots$, or $k$,
          \item or else $\beta < -2k-1$ and $-2l-1 < \alpha < -2l$ for $l=0, 1, \ldots$, or $k$,
          \item or else $-2l-2 < \alpha < -2l-1$ for $l=0$, 1, \ldots, or $k-1$ and $-2m-3 < \beta < -2m-2$ for
            $m=k-l-1, k-l, \ldots$, or $k-1$,
          \item or else $-2l-3 < \alpha < -2l-2$ for $l=0, 1, \ldots$, or $k-1$ and $-2m-2 < \beta < -2m-1$ for
            $m=k-l-1, k-l, \ldots$, or $k-1$.
        \end{itemize}
    \end{itemize}
\end{itemize}

For $n=1$, for instance, we obtain $\alpha+\beta \ne -2$ and
\begin{itemize}\itemsep=0pt
  \item [a)] $\alpha \ge 0$ and $\beta < -1$,
  \item [b)] $\alpha < -1$ and $\beta \ge 0$,
  \item [c)] $\alpha < -1$ and $-1 < \beta < 0$ or else $-1 < \alpha < 0$ and $\beta < -1$,
\end{itemize}
while for $n=2$, we get $\alpha+\beta \ne -3$, $-4$ and
\begin{itemize}\itemsep=0pt
  \item [a)] $\alpha \ge 0$ and $\beta < -2$,
  \item [b)] $\alpha < -2$ and $\beta \ge 0$,
  \item [c)] $\alpha < -2$ and $\beta < -2$ or else $\alpha < -2$ and  $-1 < \beta < 0$ or else $-1 < \alpha < 0$
    and $\beta < -2$ or else $-2 < \alpha < -1$ and $-2 < \beta < -1$.
\end{itemize}

It is worth noting that in Cases~a and~b, as well as in Case~c with $n=2k$, $\alpha < -2k$, and $\beta < -2k$, both parameters $\alpha$ and $\beta$ may take values in some extended ranges. In contrast, for the other possibilities of Case~c, at least one of the parameters $\alpha$, $\beta$ is restricted to a small interval and for this reason such subcases will be termed ``exceptional''.

On the other hand, the number of zeros of $P_n^{(\alpha, \beta)}(x)$ in $1 < x < \infty$ is given by~\cite{szego}
\begin{gather}
  N_3(\alpha, \beta) = \begin{cases}
    2 \left[\dfrac{Z+1}{2}\right] &\text{if $\displaystyle \binom{2n+\alpha+\beta}{n} \binom{n+\alpha}{n} > 0$}, \vspace{1mm}\\
    2 \left[\dfrac{Z}{2}\right] + 1 &\text{if $\displaystyle \binom{2n+\alpha+\beta}{n} \binom{n+\alpha}{n} < 0$},
  \end{cases}   \label{eq:zeros-5}
\end{gather}
where $Z$ is def\/ined by
\begin{gather}
  Z = Z(\alpha, \beta) = E \bigl[\tfrac{1}{2} (-|2n+\alpha+\beta+1| - |\alpha| + |\beta| + 1)\bigr],  \label{eq:zeros-6}
\end{gather}
with $E(u)$ given in (\ref{eq:zeros-3}). It follows that the necessary and suf\/f\/icient conditions for having no zero in $(1, \infty)$ are
\begin{gather*}
  Z = 0 \qquad \text{and} \qquad \binom{2n+\alpha+\beta}{n} \binom{n+\alpha}{n} > 0.
\end{gather*}

Such conditions can also be reformulated in an appropriate way:

\medskip\noindent {\bf Rule 2.} $P_n^{(\alpha,\beta)}(x)$ has no zero in $(1, \infty)$ if and only if $\beta \ne -1, -2, \ldots, -n$ and one of the following cases occurs:
\begin{itemize}\itemsep=0pt
  \item [a)] $\alpha < -n$ and $\alpha+\beta < -2n$,
  \item [b)] $\alpha > -1$ and $\alpha+\beta > -n-1$,
  \item [c)]
    \begin{itemize}\itemsep=0pt
      \item [(i)] for $n=2k$:
        \begin{itemize}\itemsep=0pt
          \item $\alpha < -2k$ and $-2k-2l-1 < \alpha+\beta < -2k-2l$ for $l=1, 2, \ldots$, or $k-1$ or $-2k-1 <
            \alpha+\beta$,
          \item or else $-2l-2 < \alpha < -2l-1$ for $l=0, 1, \ldots$, or $k-1$ and $-2k-2m-2 < \alpha+\beta < -2k-2m-1$
            for $m=0, 1, \ldots$, or $l$,
          \item or else $-2l-3 < \alpha < -2l-2$ for $l=0, 1, \ldots$, or $k-2$ and $-2k-2m-3 < \alpha+\beta < -2k-2m-2
            $ for $m=0, 1, \ldots$, or $l$, or $-2k-1 < \alpha+\beta$,
        \end{itemize}
      \item[(ii)] for $n=2k+1$:
        \begin{itemize}\itemsep=0pt
          \item $\alpha < -2k-1$ and $-2k-2l-1 < \alpha+\beta < -2k-2l$ for $l=1, 2, \ldots$, or $k$,
          \item or else $-2l-2 < \alpha < -2l-1$ for $l=0, 1, \ldots$, or $k-1$ and $-2k-2m-3 < \alpha+\beta < -2k-2m-2
            $ for $m=0, 1, \ldots$, or $l$,
          \item or else $-2l-3 < \alpha < -2l-2$ for $l=0, 1, \ldots$, or $k-1$ and $-2k-2m-4 < \alpha+\beta < -2k-2m-3
            $ for $m=0, 1, \ldots$, or $l$ or $-2k-2 < \alpha+\beta$.
        \end{itemize}
    \end{itemize}
\end{itemize}

For $n=1$, for instance, this leads to $\beta \ne -1$ and
\begin{itemize}\itemsep=0pt
  \item [a)] $\alpha < -1$ and $\alpha+\beta < -2$,
  \item [b)] $\alpha > -1$ and $\alpha+\beta > -2$,
\end{itemize}
while for $n=2$, we get $\beta \ne -1$, $-2$ and
\begin{itemize}\itemsep=0pt
  \item [a)] $\alpha < -2$ and $\alpha+\beta < -4$,
  \item [b)] $\alpha > -1$ and $\alpha+\beta > -3$,
  \item [c)] $\alpha < -2$ and $\alpha+\beta > -3$ or else $-2 < \alpha < -1$ and  $-4 < \alpha+\beta < -3$.
\end{itemize}

As for Rule~1, we may distinguish here Cases~a and~b, as well as Case~c with $n=2k$, $\alpha < -2k$, and $\alpha+\beta > -2k-1$, from the remaining possibilities of Case~c, which will be quoted as ``exceptional'' again.

\vspace{-2mm}

\subsection*{Acknowledgements}

The author would like to thank Y.~Grandati for several useful discussions.

\vspace{-3mm}

\pdfbookmark[1]{References}{ref}
\LastPageEnding

\end{document}